\documentclass[prx,twocolumn,superscriptaddress]{revtex4}

\usepackage{graphicx,float}
\usepackage{braket}
\usepackage{amsmath}
\usepackage{amsfonts}
\usepackage{xcolor}
\usepackage{todonotes}
\usepackage{hhline}
\usepackage{euscript}
\usepackage{siunitx}
\usepackage{changes}
\DeclareMathOperator{\arccosh}{arccosh}

\begin{document}

\title{Engineering a Kerr-based Deterministic Cubic Phase Gate via Gaussian Operations}

\author{Ryotatsu Yanagimoto}
\thanks{These authors contributed equally to this work. \\Email: ryotatsu@stanford.edu, \\{\color{white}Email: }tatsuhiro.onodera@ntt-research.com}
\affiliation{E.\,L. Ginzton Laboratory, Stanford University, Stanford, California 94305, USA}

\author{Tatsuhiro Onodera}
\thanks{These authors contributed equally to this work. \\Email: ryotatsu@stanford.edu, \\{\color{white}Email: }tatsuhiro.onodera@ntt-research.com}
\affiliation{E.\,L. Ginzton Laboratory, Stanford University, Stanford, California 94305, USA}
\affiliation{NTT Physics and Informatics Laboratories, NTT Research, Inc., 1950 University Ave. East Palo Alto, CA 94303, USA}
\affiliation{School of Applied and Engineering Physics, Cornell University, Ithaca, New York 14853, USA}

\author{Edwin Ng}
\affiliation{E.\,L. Ginzton Laboratory, Stanford University, Stanford, California 94305, USA}

 \author{Logan G. Wright}
\affiliation{E.\,L. Ginzton Laboratory, Stanford University, Stanford, California 94305, USA}
\affiliation{NTT Physics and Informatics Laboratories, NTT Research, Inc., 1950 University Ave. East Palo Alto, CA 94303, USA}
\affiliation{School of Applied and Engineering Physics, Cornell University, Ithaca, New York 14853, USA}

\author{Peter L. McMahon}
\affiliation{School of Applied and Engineering Physics, Cornell University, Ithaca, New York 14853, USA}

\author{Hideo Mabuchi}
\affiliation{E.\,L. Ginzton Laboratory, Stanford University, Stanford, California 94305, USA}

\begin{abstract}
We propose a deterministic, measurement-free implementation of a cubic phase gate for continuous-variable quantum information processing. In our scheme, the applications of displacement and squeezing operations allow us to engineer the effective evolution of the quantum state propagating through an optical Kerr nonlinearity. Under appropriate conditions, we show that the input state evolves according to a cubic phase Hamiltonian, and we find that the cubic phase gate error decreases inverse-quartically with the amount of quadrature squeezing, even in the presence of linear loss. We also show how our scheme can be adapted to deterministically generate a nonclassical approximate cubic phase state with high fidelity using a ratio of native nonlinearity to linear loss of only \num{e-4}, indicating that our approach may be experimentally viable in the near term even on all-optical platforms, e.g., using quantum solitons in pulsed nonlinear nanophotonics.
\end{abstract}

\date{\today}

\maketitle
Quantum computation (QC) holds promise for outperforming conventional computers in solving certain types of hard problems of practical interest~\cite{Nielsen2000}. Alongside established discrete-variable QC platforms such as superconducting microwave circuits~\cite{DiCarlo2009,Plantenberg2007} and trapped ions~\cite{Leibfried2003}, continuous-variable (CV) quantum information processing using optical systems is also expected to play important roles in the path towards practical QC~\cite{Gottesman2001,Braunstein2005,Menicucci2006,Kimble2008,Weedbrook2012,Takeda2017,Takeda2019,Asavanant2019}, due to its room-temperature and high-bandwidth operability. In such systems, it is known that Gaussian operations together with one non-Gaussian operation suffices to realize a universal gate set, and in many proposed CVQC architectures, the cubic phase gate $\hat{U}^\mathrm{cubic}(\gamma)=\exp(\mathrm{i}\gamma\hat{x}^3)$ is a particularly convenient candidate for the non-Gaussian operation~\cite{Gottesman2001,Braunstein2005,Gu2009}. A cubic phase gate can be implemented by utilizing the nonlinearity induced by non-Gaussian photon measurement together with Gaussian feed-forward operations~\cite{Knill2001, Marshall2015,Marek2011,Yukawa2013}. Alternatively, Gaussian measurement and feed-forward operations suffice if a cubic phase state $\ket{\gamma}=\int^\infty_{-\infty}e^{\mathrm{i}\gamma x^3}\ket{x}\mathrm{d}x$ is available as an ancilla~\cite{Gottesman2001,Bartlett2002,Marek2018}, and approximate constructions of this state under photon-number constraints are known~\cite{Miyata2016}. 

To our knowledge, however, no measurement-free (i.e., \emph{coherent}) implementation of a cubic phase gate has yet been proposed. This is a challenging task because (i) optical material nonlinearities tend to be small relative to typical dissipation rates, and (ii) it is difficult to find a simple natural process described by a cubic-phase Hamiltonian $\hat{H}^\text{cubic}\propto\hat{x}^3$. To address these issues, we take inspiration from other proposals where Gaussian operations are utilized to alter and enhance the quantum dynamics of a system. For example, it is known that by selectively and strongly displacing one of the optical fields in a native four-wave mixing process, one can control how the photons in the other fields interact~\cite{Langford2011,Ramelow2019}, while the use of squeezed light can enhance coupling of the optical field to other degrees of freedom~\cite{Qin2018,Leroux2018,Michael2019}. In this work, we apply both squeezing and displacement operations on the state before entering a Kerr nonlinearity---and then apply the inverse of those operations after---to realize $\hat U^\text{cubic}$ coherently. Intuitively, the squeezing selectively enhances the sensitivity of the $x$-quadrature to the Kerr effect, effectively forming in phase space a 1D quartic potential in $x$; the displacement then allows us to choose an operating point along $x$ in the vicinity of which the potential is approximately cubic in $x$.

\begin{figure}[t]
    \includegraphics[width=0.45\textwidth]{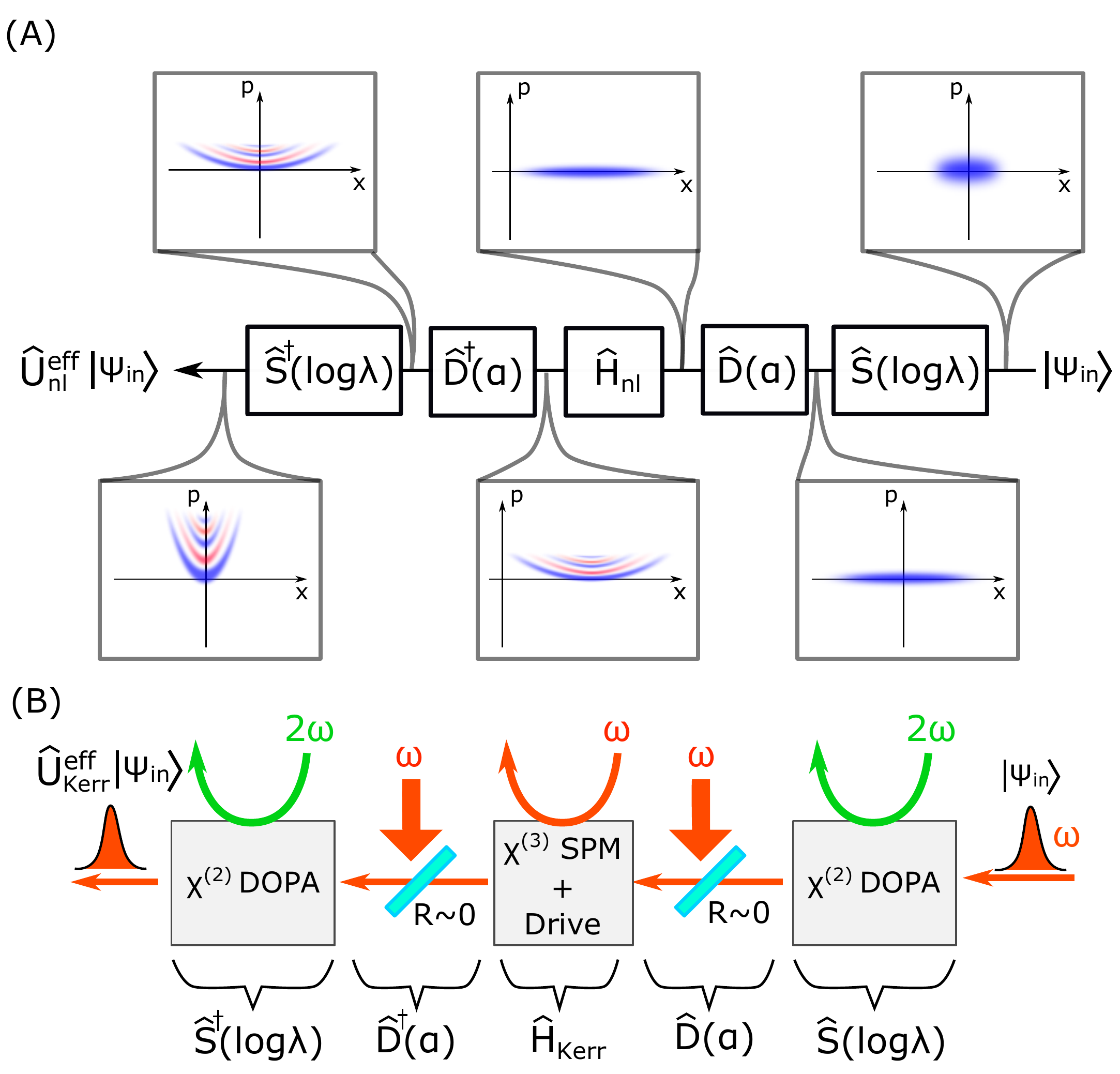}
    \caption{(A) Gaussian operations are applied to the input state $\ket{\psi_\mathrm{in}}$ before and after evolution under the Hamiltonian $\hat{H}_\mathrm{nl}$. Insets: Wigner functions of the quantum state at each step of generating a cubic phase state. (B) Possible realization of a coherent cubic phase gate acting on a pulse with carrier frequency $\omega$. Here, $\hat{H}_\mathrm{Kerr}$ is realized by $\chi^{(3)}$ self-phase modulation (SPM) in combination with a detuned drive. Displacement and squeezing can be realized using high-transmission beamsplitters and $\chi^{(2)}$ degenerate parametric amplifiers (DOPAs) pumped at $2\omega$, respectively.}
    \label{fig:unitary}
\end{figure}

The general setup of our scheme is as follows. We consider a nonlinear medium with some physical native Hamiltonian $\hat H_\mathrm{nl} = \sum_{n,m} c_{n,m} (\hat{a}^\dagger)^n (\hat{a})^m$, where $\hat{a}$ is a mode annihilation operator. As shown in Fig.~\ref{fig:unitary}(A), before the input state $\ket{\psi_\mathrm{in}}$ enters the nonlinear medium, we first apply two operations: a squeezing operation $\hat{S}(\log\lambda)$, corresponding to in-phase quadrature power gain $\lambda^2$ and a displacement operation $\hat{D}(\alpha)$. Then, after propagation through the nonlinear medium for time $\tau$, we apply the inverse operations $\hat{D}^\dagger(\alpha)$ and $\hat{S}^\dagger(\log\lambda)$. This sequence can be described by a unitary operator
\begin{align}
    \hat{U}^\mathrm{eff}_\mathrm{nl}=\hat{S}^\dagger(\log\lambda)\hat{D}^\dagger(\alpha)e^{-\mathrm{i}\hat{H}_\mathrm{nl}\tau}\hat{D}(\alpha)\hat{S}(\log\lambda).
\end{align}
This operator can be associated with an effective Hamiltonian $\hat H_\text{nl}^\text{eff}$ such that $\hat{U}^\mathrm{eff}_\mathrm{nl} = e^{-\mathrm{i}\hat{H}_\mathrm{nl}^\mathrm{eff}\tau}$, where $\hat{H}_\mathrm{nl}^\mathrm{eff}$ is given by $\hat H_\text{nl}$ under the substitution $\hat{a} \mapsto \hat{a}_\mathrm{eff}=\lambda\hat{x}/\sqrt{2}+\mathrm{i}\lambda^{-1}\hat{p}/\sqrt{2}+\alpha$; explicitly,
\begin{align}
\begin{split}
   \hat H_\mathrm{nl}^\mathrm{eff}&=\frac{1}{2}\sum_{n,m}c_{n,m}\lambda^{2(n+m)}(\hat{x}-\mathrm{i}\lambda^{-2}\hat{p}+\sqrt{2}\alpha)^n \\
    &\quad\times(\hat{x}+\mathrm{i}\lambda^{-2}\hat{p}+\sqrt{2}\alpha)^m.
    \end{split}
    \label{eq:Heff}
\end{align}

In the limit of strong squeezing $\lambda\rightarrow\infty$, this Hamiltonian converges to a polynomial of $\hat{x}$, so intuitively, by properly choosing $c_{n,m}$ and $\alpha$ in this limit, one can engineer a variety of $\hat x$-dependent effective Hamiltonians, such as the cubic phase Hamiltonian. In addition, by comparing the terms in $\hat H_\text{nl}$ with those of the same order in $\hat H_\text{nl}^\text{eff}$, we see the inclusion of $\lambda$ and $\alpha$ can lead to a \emph{Gaussian enhancement} of the native Hamiltonian coefficients. Though we focus on the explicit construction and evaluation of a cubic phase gate in this paper, it is worth mentioning that this approach can in principle be extended to a much broader class of gates.

More concretely, consider a driven Kerr Hamiltonian
\begin{align}
    \label{eq:kerr}
\hat{H}_\mathrm{Kerr}=-\frac{\chi}{2}\hat{a}^\dagger{}^2\hat{a}^2+\delta\hat{a}^\dagger\hat{a}+\beta(\hat{a}+\hat{a}^\dagger).
\end{align}
A possible construction of such a Hamiltonian is shown schematically in Fig.~\ref{fig:unitary}(B), where $\chi$ is associated with the $\chi^{(3)}$ self-phase modulation of a propagating pulse in a Kerr medium. Alternatively, it can also be realized in a driven high-finesse Kerr cavity~\cite{Heuck2019,Heuck2019b}. In some settings, the application of a constant drive concurrent with the nonlinearity may be technologically challenging; in Appendix~\ref{sec:trotter}, we show an alternative realization by applying the drive and nonlinearity separately, in the form of additional displacements and undriven Kerr mediums.

We now apply the transformation $\hat a \mapsto \hat a_\text{eff}$ to \eqref{eq:kerr}, assuming $\alpha>0$. After some cumbersome but straightforward calculations, we get the effective Hamiltonian
\begin{widetext}
    \begin{equation}
    \label{eq:Heffkerr}
        \begin{split}
        \hat{H}^\mathrm{eff}_\mathrm{Kerr}(\delta,\beta)=&-\frac{\chi}{8}(\lambda^4\hat{x}^4
        +\hat{p}\hat{x}^2\hat{p}+\hat{x}\hat{p}^2\hat{x}+\lambda^{-4}\hat{p}^4)-\frac{1}{\sqrt{2}}\chi\lambda^3\alpha\hat{x}^3-\frac{1}{\sqrt{2}}\chi\lambda^{-1}\alpha\hat{p}\hat{x}\hat{p}\\
        &+\frac{1}{2}\lambda^2\left(-3\chi\alpha^2+\chi+\delta\right)\hat{x}^2+\frac{1}{2}\lambda^{-2}(-\chi\alpha^2+\chi+\delta)\hat{p}^2+\sqrt{2}\lambda(-\chi\alpha^3+\chi\alpha+\delta\alpha+\beta)\hat{x},
        \end{split}
        \end{equation}
\end{widetext}
where we have ignored a constant offset. Notice that the cubic terms in $\hat x$ and $\hat p$ carry a factor of $\alpha$: intuitively, these terms arise when considering how the quantum fluctuations of three of the fields in the four-wave-mixing process interact in the presence of the mean of the fourth; i.e., when the latter is formally replaced by its classical displacement $\alpha$.

Now, we attempt to cancel undesired terms by an appropriate choice of the detuning $\delta^\mathrm{cubic}$ and drive $\beta^\mathrm{cubic}$ to arrive at an effective cubic phase Hamiltonian $\hat{H}^\mathrm{cubic}_\mathrm{Kerr} = \hat{H}^\mathrm{eff}_\mathrm{Kerr}(\delta^\mathrm{cubic},\beta^\mathrm{cubic})$ which approximates the desired form $\sim \hat{x}^3$. Specifically, we set $\delta^\mathrm{cubic}=3\chi\alpha^2-\chi$ and $\beta^\mathrm{cubic}=-2\chi\alpha^3$ to eliminate the $\hat{x}^2$ and $\hat{x}$ terms, and then assume a scaling $\alpha \sim \lambda^3$.  This produces a Kerr-based cubic phase Hamiltonian
\begin{small}
\begin{align}
    \label{eq:lowestorder}
    \hat{H}^\mathrm{cubic}_\mathrm{Kerr}=-\mu\left(\hat{x}^3+\frac{\lambda\alpha^{-1}}{4\sqrt{2}}\hat{x}^4-\sqrt{2}\lambda^{-5}\alpha\hat{p}^2+\mathcal{O}(\lambda^{-4})\right)
\end{align}
\end{small}where $\mu=\frac{\chi\lambda^3\alpha}{\sqrt{2}}>0$.  Intuitively, the scaling of $\alpha \sim \lambda^3$ is assumed to minimize the second and third terms above, and we later numerically show that this scaling is optimal. Thus, to leading order in $1/\lambda$, the Hamiltonian takes the desired form, and a cubic phase gate $\hat{U}^\mathrm{cubic}_\mathrm{Kerr}(\gamma)=\exp(-\mathrm{i}\tau \hat{H}^\mathrm{cubic}_\mathrm{Kerr})$ with gate angle $\gamma$ is realized by setting the gate time to be $\tau=\sqrt{2}\gamma/\chi\alpha\lambda^3$.

Because the amplitude of the major error terms scales as $\mathcal{O}(\lambda^{-2})$, there is an intrinsic gate error even in the absence of any decoherence. We can define this error to be $\mathcal E_\text{int} = \mathcal E(\hat U_\text{Kerr}^\text{cubic}\ket{\psi_\text{in}}) = 1 -\mathcal{F}(\hat U_\text{Kerr}^\text{cubic}\ket{\psi_\text{in}},\hat{U}^\text{cubic}\ket{\psi_{\mathrm{in}}})$, where $\mathcal F$ is the fidelity. Based on the scaling of the error terms, we expect $\mathcal E_\text{int}$ to scale as $\mathcal{O}(\lambda^{-4})$.

To concretely evaluate our scheme, we consider the Gottesman-Kitaev-Preskill (GKP) qubit states~\cite{Gottesman2001,Albert2018,Fuehmann2109} as inputs and evaluate the gate error for $\gamma=0.1$. The symmetric GKP qubits that we consider are
\begin{small}
\begin{align}
\label{eq:gkp}
&\ket{z^+}\propto\sum^\infty_{k=-\infty}\exp\left[-\frac{(2k\sqrt{\pi}\Delta)^2}{2}\right]\hat{D}(2k\sqrt{\pi})\ket{\Delta} \\
&\ket{z^{-}}\propto\sum^\infty_{k=-\infty}\exp\left[-\frac{\big((2k+1)\sqrt{\pi}\Delta\big)^2}{2}\right]\hat{D}\big((2k+1)\sqrt{\pi}\big)\ket{\Delta}, \nonumber
\end{align}
\end{small}where $\ket{\Delta}$ is a squeezed vacuum state with variance $\langle\hat{p}^2\rangle-\langle\hat{p}\rangle^2=\Delta^2/2$. We refer to the other relevant states in the qubit subspace as $\ket{y^\pm}=(\ket{z^+}\pm\mathrm{i}\ket{z^{-}})/\sqrt{2}$ and $\ket{x^\pm}=(\ket{z^+}\pm\ket{z^{-}})/\sqrt{2}$.
\begin{figure}[b!]
    \includegraphics[width=0.46\textwidth]{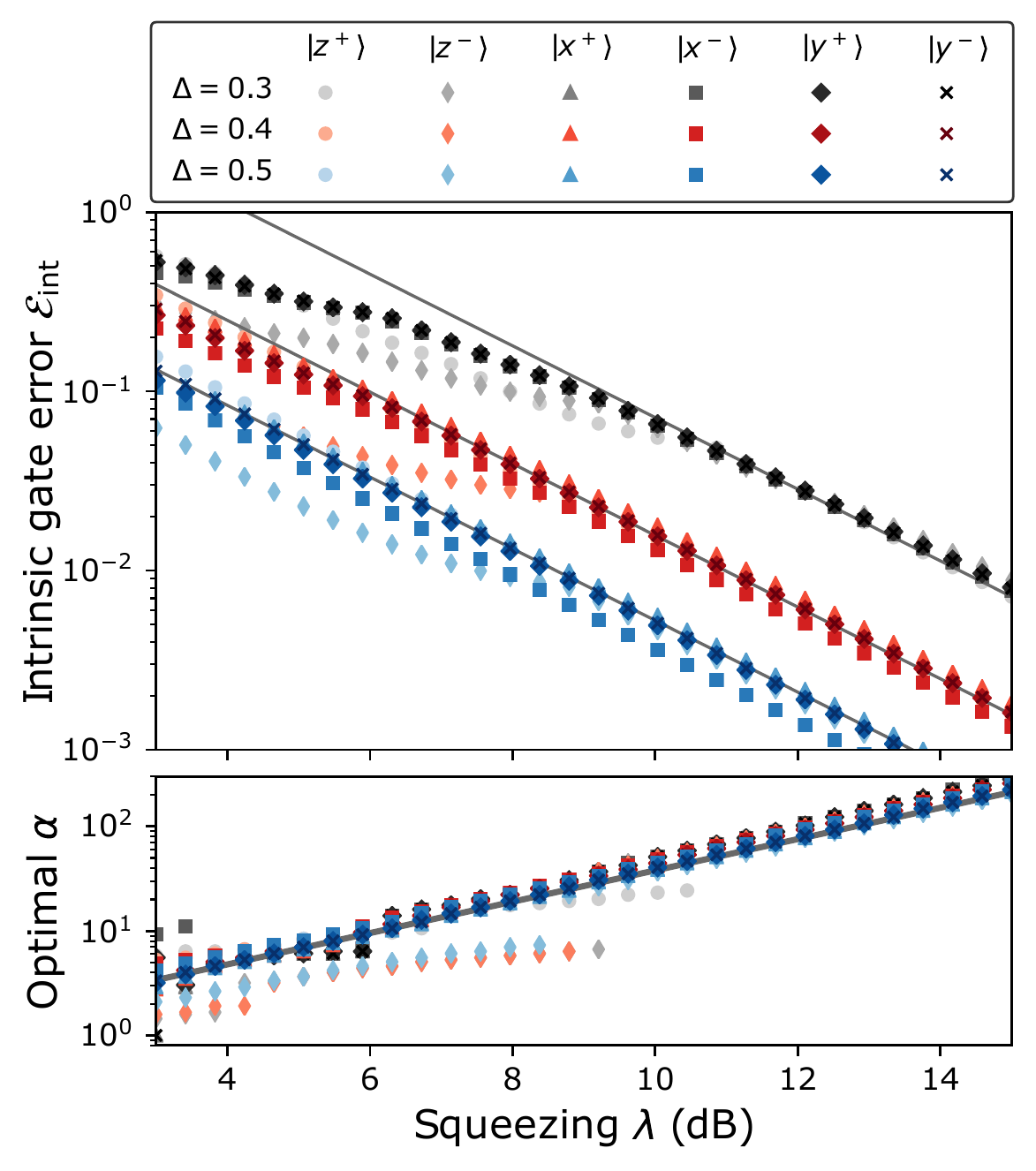}
    \caption{Intrinsic gate error $\mathcal{E}_\mathrm{int}$ (upper) and optimal displacement $\alpha$ (lower) of the proposed coherent cubic phase gate with gate angle $\gamma=0.1$ for various GKP qubit states \eqref{eq:gkp} with three values of $\Delta$, as a function of the squeezing $\lambda$. The grey solid lines are guides that show the predicted scaling $\mathcal{E}_\mathrm{int}\sim\lambda^{-4}$ (upper) and $\alpha\sim\lambda^3$ (lower).}
    \label{fig:fidelitynoloss}
\end{figure}

We show the intrinsic gate errors for GKP qubit states with $\Delta=0.3$, $0.4$, and $0.5$ in Fig.~\ref{fig:fidelitynoloss}, where we observe the expected scaling $\mathcal{E}_\mathrm{int}\sim\lambda^{-4}$ holds for all the relevant GKP qubit states for large enough $\lambda$. In this calculation, we optimized over the displacement $\alpha$ to minimize the gate error, and we indeed observe the theoretically expected scaling $\alpha=C\lambda^3$. As shown in the figure, we also find that the constant $C$ does not depend strongly on the input state. Based on these observations, we choose to focus our analysis of the gate performance on the input state $\ket{z^+}$ with $\Delta=0.5$ without much loss of generality.

In considering the implementation of our scheme on optical platforms, the actual limiting factor will most likely be the linear loss, and here we characterize its impact. Specifically, we focus on linear loss that acts concurrently with the Kerr nonlinearity; we model this loss using a Lindblad operator $\hat{L}=\sqrt{\kappa}\hat{a}$, where $\kappa$ is the power decay rate. Due to the Gaussian transformations, the effective Lindblad operator is $\sqrt\kappa \hat a_\text{eff}$. Thus, though the native linear loss rate is $\kappa$, the effective loss rate is $\sim\lambda^2\kappa$, corresponding to increased quantum fluctuations. On the other hand, the time required to realize a desired gate operation is $\tau\sim\lambda^{-6}$. As a result, the gate error induced by the linear loss should scale as $\kappa\tau\sim \lambda^{-4}$. Thus, the total error $\mathcal{E} = \mathcal E(\hat\rho)$, where $\hat{\rho}$ is the output state with linear loss, also scales inverse-quartically with respect to the field gain $\lambda$ of the squeezer.

\begin{figure}
    \includegraphics[width=0.5\textwidth]{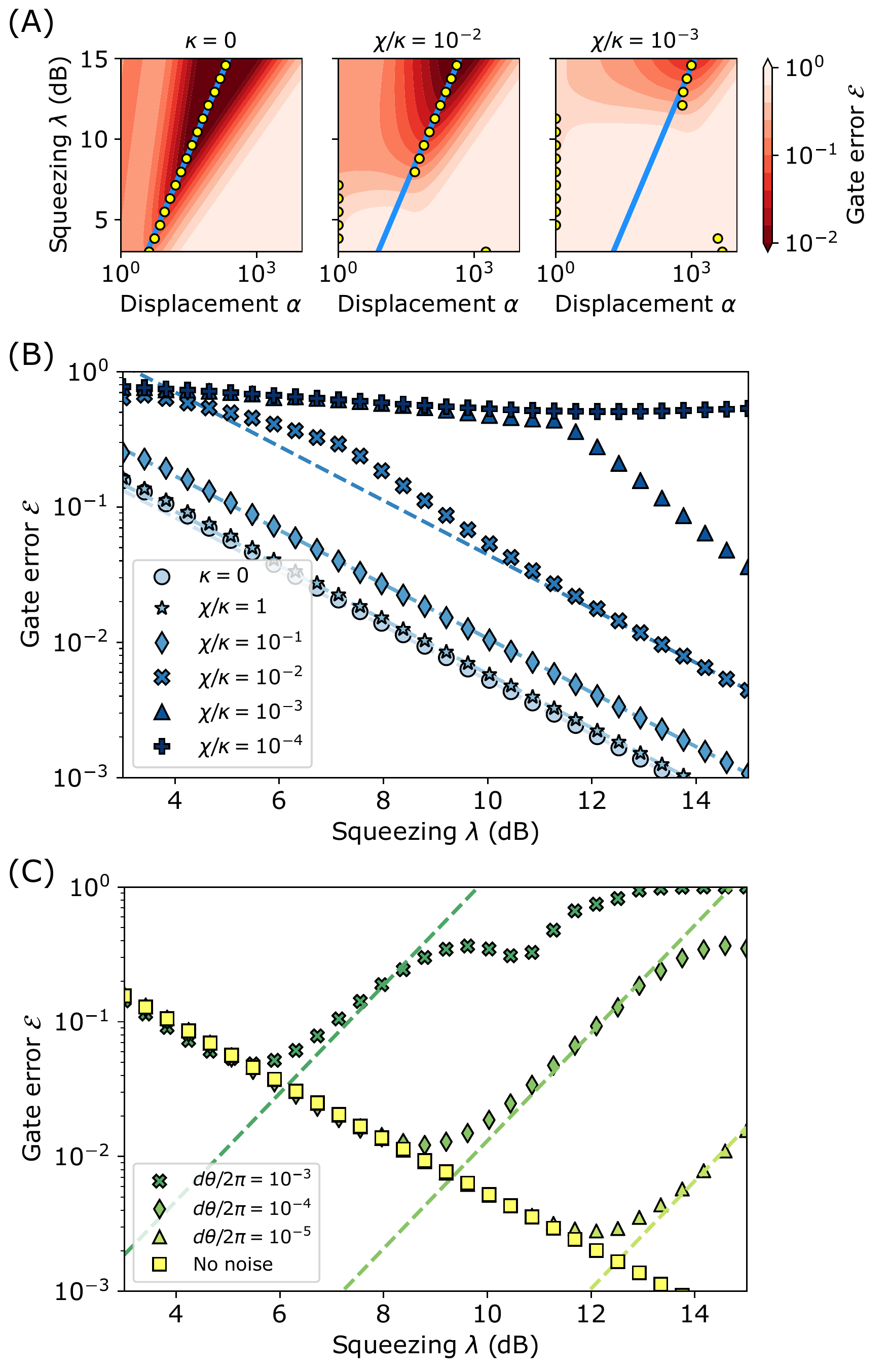}
    \caption{(A) Gate error $\mathcal E$ as a function of displacement $\alpha$ and squeezing $\lambda$ for three different values of linear loss $\kappa$. Yellow circles show optimal $\alpha$ for each $\lambda$; blue solid lines show the expected scaling $\alpha\propto\lambda^3$. (B) Gate error as a function of $\lambda$ for the optimal choice of $\alpha$. Dashed lines show the expected scaling of $\mathcal{E}\sim\lambda^{-4}$. (C) Gate error in the presence of a small phase rotation $\mathrm{d}\theta$ that is applied after the propagation through the nonlinear medium. Dashed lines show the expected scaling $\mathcal{E}\sim\lambda^8$. For all the plots, the input state is the GKP qubit state \eqref{eq:gkp} $\ket{z^{+}}$ with $\Delta=0.5$, and $\gamma=0.1$.}
    \label{fig:gkp_combined}
\end{figure}

In order to confirm these heuristic scaling arguments, we perform numerical simulations of the gate operations in the presence of linear loss. First, in Fig.~\ref{fig:gkp_combined}(A), we show the gate error for various values of $\kappa$ as a function of $\lambda$ and $\alpha$. The figure shows that the optimal $\alpha$ that minimizes the gate error for a given $\lambda$ increases for larger $\kappa$. This is because larger $\alpha$ reduces the gate time $\tau=\sqrt{2}\gamma/\chi\alpha\lambda^3$, which decreases the error incurred by linear loss, while an $\alpha$ that is too large leads to an increase in $\mathcal{E}_\mathrm{int}$. Thus, the optimal $\alpha$ is determined based on the balance between these two contributions, and when linear loss dominates, larger $\alpha$ is preferable. The gate error achieved for such optimal $\alpha$ is shown in Fig.~\ref{fig:gkp_combined}(B), where the expected scaling of $\mathcal{E}\sim\lambda^{-4}$ is observed.

Next, we analyze the robustness of our scheme by considering the effect of stochastic variations of system parameters on gate error. Because of the Gaussian enhancement in our scheme, the effects of these variations could be more pronounced. Here, we investigate phase noise, which might arise, e.g., from phase deviations between the input and output squeezers. We model the noise as a fixed phase rotation $\mathrm{d}\theta/2\pi$ applied at the end of the nonlinear medium. In the presence of this noise, the unitary is
\begin{small}\begin{align}
   \exp(-\mathrm{id}\theta\hat{a}_\mathrm{eff}^\dagger\hat{a}_\mathrm{eff})\hat{U}^\mathrm{cubic}_\mathrm{Kerr}\sim \exp\left(-\sqrt{2}\mathrm{i}\lambda\alpha\hat{x}\mathrm{d}\theta\right)\hat{U}^\mathrm{cubic}_\mathrm{Kerr}.
\end{align}\end{small}Therefore, to leading order, a phase rotation at the end of the nonlinear medium is equivalent to a $p$-displacement on the output state by an amount proportional to $\lambda^4\mathrm{d}\theta$. The gate error due to this noise scales as $(\lambda^4\mathrm{d}\theta)^2$, as confirmed by the numerical simulation shown in Fig.~\ref{fig:gkp_combined}(C). In Appendix \ref{sec:noiseanalysis}, we show that other major sources of noise are also described by $p$-displacements with the same scaling. Conveniently, such displacement errors can be naturally corrected in the GKP scheme~\cite{Gottesman2001}. In order to implement a gate with $\mathcal{E}<1$\%, a low phase noise of $\mathrm{d}\theta/2\pi<\num{e-4}$ is required. Thus, careful phase stabilization of the light source and optical paths is essential for our scheme~\cite{Ludlow2007,Newman2019}. 

Next, we adapt the coherent cubic phase gate for state generation. Unlike the gate operations on an unknown state, additional Gaussian transformations can be applied to further improve the fidelity of state preparation as the input state is fixed for a desired output state. In particular, we focus on the generation of a cubic phase state $\ket{\gamma}$, which is the output of an ideal cubic phase gate acting on an infinitely squeezed vacuum state $\lim_{\Delta\rightarrow 0}\ket{\Delta}$. Because such states are unbounded, we instead use a finitely squeezed vacuum $\ket{\Delta=0.5}$ as an input state, which consequently sets the target state to be $\hat{U}^\text{cubic}(\gamma)\ket{\Delta=0.5}$. Following the discussion in Ref.~\cite{Miyata2016}, the discrepancy between this chosen state and $\ket{\gamma}$ is quantified by the variance of the operator $\hat{p}_\mathrm{NLQ}=\hat{p}-3\gamma\hat{x}^2$, which is $\frac{\Delta^2}{2}=0.125$.

\begin{figure}[h]
    \includegraphics[width=0.50\textwidth]{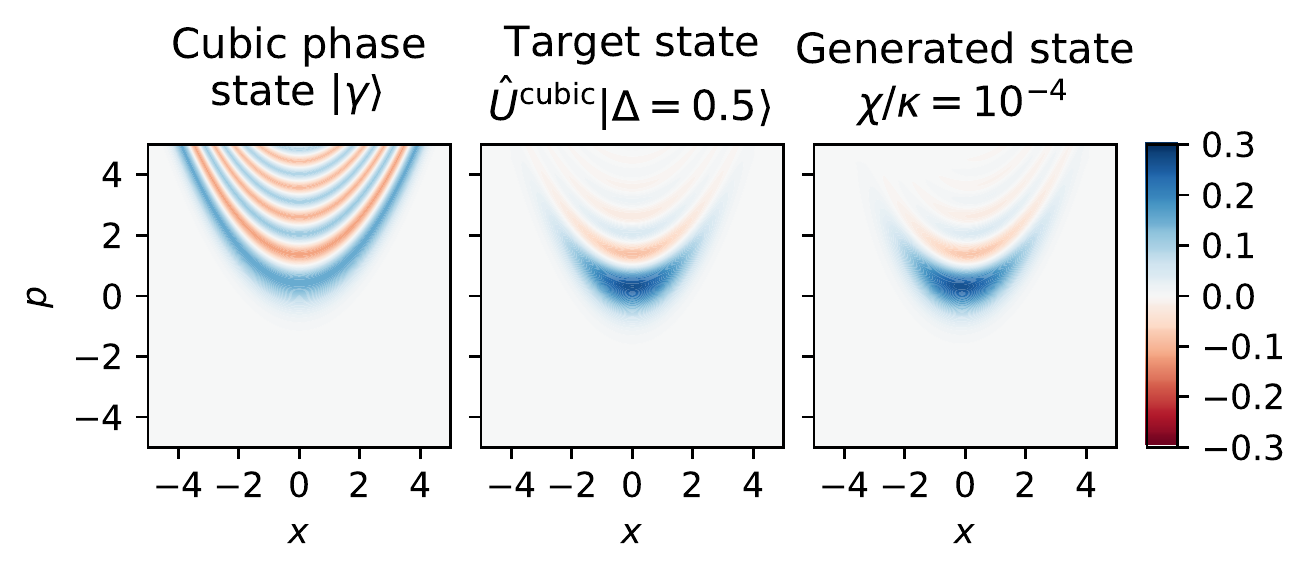}
    \caption{Left: Wigner functions of the cubic phase state with $\gamma=0.1$, which is normalized within the plotted region. Middle: An approximate cubic phase state generated by acting with the ideal cubic phase gate on the squeezed state $\ket{\Delta=0.5}$. Right: Cubic phase state generated using a coherent cubic phase gate (our scheme) acting on $\ket{\Delta=0.5}$ with linear loss $\chi/\kappa=10^{-4}$, \SI{15}{dB} of squeezing $\lambda$, and $\alpha=1.4\times 10^4$.}
    \label{fig:wignerplots}
\end{figure}

In Fig.~\ref{fig:wignerplots}, we show the Wigner function of a cubic phase state realized with squeezing $\lambda$ of \SI{15}{dB} and displacement $\alpha=1.4\times 10^4$ as an example. Even with a large linear loss of $\chi/\kappa=10^{-4}$, the characteristic oscillations of the Wigner function~\cite{Ghose2007} of an ideal cubic phase state are reproduced, and the fidelity of the state with respect to the target state is $97.8\%$. This is in stark contrast to the generation of Fock states~\cite{Yanagimoto2019} for instance, where the same Hamiltonian as \eqref{eq:kerr} is used, but $\chi/\kappa>10^3$ is needed to reach this level of fidelity. This resilience of our scheme to linear loss is attributed to the Gaussian amplification of the nonlinearity.

Though the scheme we presented is general and can be implemented with continuous-wave optics or even superconducting circuits, we briefly discuss the experimental feasibility of our scheme in the context of pulsed nonlinear optics, as time-multiplexed qubits encoded in optical pulses is a convenient platform for scalable quantum information processing~\cite{Takeda2019}. In this context, however, the time-domain nature of the nonlinear interaction can, in general, lead to spectral-temporal mode distortions of the pulse in which the state resides~\cite{Boivin1994,Joneckis1993,Schmidt1998,Blow1991,Leung2009,Brecht2011}. As we show in Appendix~\ref{sec:qsoliton}, one way to ensure single-mode behavior of an optical pulse propagating in a Kerr medium is to encode the state into a quantum Kerr soliton~\cite{Shirasaki1990,Wright1991, Lai1998,Lai1998b,Korolkova2001,Chen2007,Drummond2014}, which can be done provided the input state is highly displaced. Fortuitously, the displacement operation in our scheme prior to entering the nonlinear medium ensures that this condition is met. In this implementation, the figure of merit is the ratio between the single-mode self-phase modulation rate and the decoherence rate of a quantum Kerr soliton, which can be expressed in terms of experimentally relevant parameters as follows:
\begin{align}
\label{eq:fom}
    \frac{\chi}{\kappa}\approx\frac{3.83\hbar\omega_0\gamma_\mathrm{nl}}{\alpha_\mathrm{att} T_\mathrm{FWHM}},
\end{align} 
where $\omega_0$ is the carrier frequency, $\gamma_\mathrm{nl}$ is the effective nonlinear coefficient, $\alpha_\mathrm{att}$ is the attenuation rate (in units of \si[power-font=unit]{\dB}/[length]), and $T_\mathrm{FWHM}$ is the full width at half-maximum of the pulse intensity in time. Assuming a soliton with $T_{\mathrm{FWHM}}=\SI{100}{fs}$, experimentally-obtained values for AlGaAs-on-insulator~\cite{Pu2016} (Si-on-insulator~\cite{Clemmen2009}) waveguides with $\gamma_{\mathrm{nl}}=\SI{660}{W^{-1} m^{-1}}$ ($\SI{280}{W^{-1} m^{-1}}$) and $\alpha_{\mathrm{att}}=\SI{140}{dB/m}$ ($\SI{400}{dB/m}$) at a wavelength of $\SI{1.59}{\micro m}$ ($\SI{1.54}{\micro m}$) results in $\chi/\kappa=2.2\times 10^{-5}$ ($3.4\times10^{-6}$). Though careful dispersion engineering~\cite{Foster2008} will be required to form a soliton with the desired length while still satisfying the requirements of our scheme (see Appendix~\ref{sec:qsoliton} for an extended discussion), these numbers indicate that, by leveraging the longitudinal as well as transverse field confinement available in pulsed nonlinear waveguides, the realization of our scheme is potentially within experimental reach in the near term even in an all-optical platform.

We have proposed a deterministic, measurement-free implementation of a cubic phase gate using Kerr nonlinearities and Gaussian operations, where the gate error is shown to exhibit a favorable scaling of $\mathcal{E}\sim\lambda^{-4}$ with respect to the quadrature squeezing $\lambda$, even in the presence of linear loss. While not specific to optics, our approach presents a potential solution, complementary to proposals based on measurement feedback~\cite{Knill2001,Bartlett2002,Kok2007,Ra2019}, to the central challenge in optical information processing of overcoming the lack of non-Gaussian (non-Clifford) gates. Despite the weak material nonlinearities in optics, the enhancement of the nonlinearity achieved by Gaussian transformations (the displacement and squeezing of the field) makes the scheme plausibly realizable when combined with high electric field confinement of optical pulses inside nanophotonic waveguides. While our scheme has immediate implications for optical QC, we also believe our technique of using Gaussian transformations to realize and enhance desired quantum operations will find broad application in optical quantum information processing and quantum engineering in general.

\begin{acknowledgments}
This work has been supported by the Army Research Office under Grant No. W911NF-16-1-0086, NTT PHI Laboratory of NTT Research, and by a Google Faculty Research Award. RY, TO, and EN are supported by National Science Foundation under award PHY-1648807. RY is also supported by Masason Foundation. LGW acknowledges additional support from Cornell Neurotech. The authors would like to thank Marc Jankowski for helpful discussions.
\end{acknowledgments}
\onecolumngrid

\appendix
\makeatletter
\section{Coherent cubic phase gate without continuous linear drive}
\label{sec:trotter}
In the main text, our scheme is based on a Kerr nonlinear Hamiltonian \eqref{eq:kerr} with a continuous linear drive term $\beta(\hat{a}+\hat{a}^\dagger)$. In this section, we introduce an alternative construction to approximately realize \eqref{eq:kerr} by applying discrete displacements before and after propagation through an undriven Kerr medium $N$ times. We justify this scheme via a simple analytic argument and run numerical simulations to verify its performance. The unitary operator that describes this sequence of operations is given by 
\begin{align}
\hat U_{\mathrm{discrete}} = \left[\hat{D}\left(-\frac{\mathrm{i}\tau\beta}{2N}\right)\exp\left(-\mathrm{i}\frac{\tau}{N}\hat{H}_\mathrm{Kerr}(\delta,\beta=0)\right)\hat{D}\left(-\frac{\mathrm{i}\tau\beta}{2N}\right)\right]^N,
\end{align}
where $\hat{D}\left(-\frac{\mathrm{i}\tau\beta}{2N}\right) = \exp\left(-\frac{\mathrm{i}\tau}{2N}\beta(\hat{a}+\hat{a}^\dagger)\right)$ are discrete displacement operations that are applied before and after propagation through the Kerr medium to approximate the drive term. Using the Trotter-Suzuki expansion, we formally see that $\hat U_{\mathrm{discrete}} = \exp\left(-\mathrm{i}\tau\hat{H}_\mathrm{Kerr}(\delta,\beta)\right)+\mathcal{O}\left(\frac{\tau^3}{N^2}\right)$. Now by applying our usual Gaussian operations before and after this entire sequence of operations, we get
\begin{align}
\hat{U}_\mathrm{discrete}^\mathrm{cubic}(\gamma,N)= \hat{S}^\dagger(\log \lambda) \hat{D}^\dagger(\alpha) \hat U_{\mathrm{discrete}} \hat{S}(\log \lambda) \hat{D}(\alpha) =\hat{U}_\mathrm{Kerr}^\mathrm{cubic}(\gamma)+\mathcal{O}\left(\frac{\gamma^3}{N^2}\right),
\end{align}
where we set $\delta=\delta^\mathrm{cubic}$, $\beta=\beta^\mathrm{cubic}$, $\tau=\frac{\sqrt{2}\gamma}{\chi\alpha\lambda^3}$ to realize the cubic phase gate. It should be noted that for $N=1$, the discrete displacements $\hat{D}\left(-\frac{\mathrm{i}\tau\beta}{2N}\right)$ can be merged with the initial displacement operation $\hat{D}(\alpha)$ of our scheme so that this alternative discretized approach at $N = 1$ can be realized in the originally proposed experimental setup without a continuous drive. 

\begin{figure}[h]
    \centering
    \includegraphics[width=0.55\textwidth]{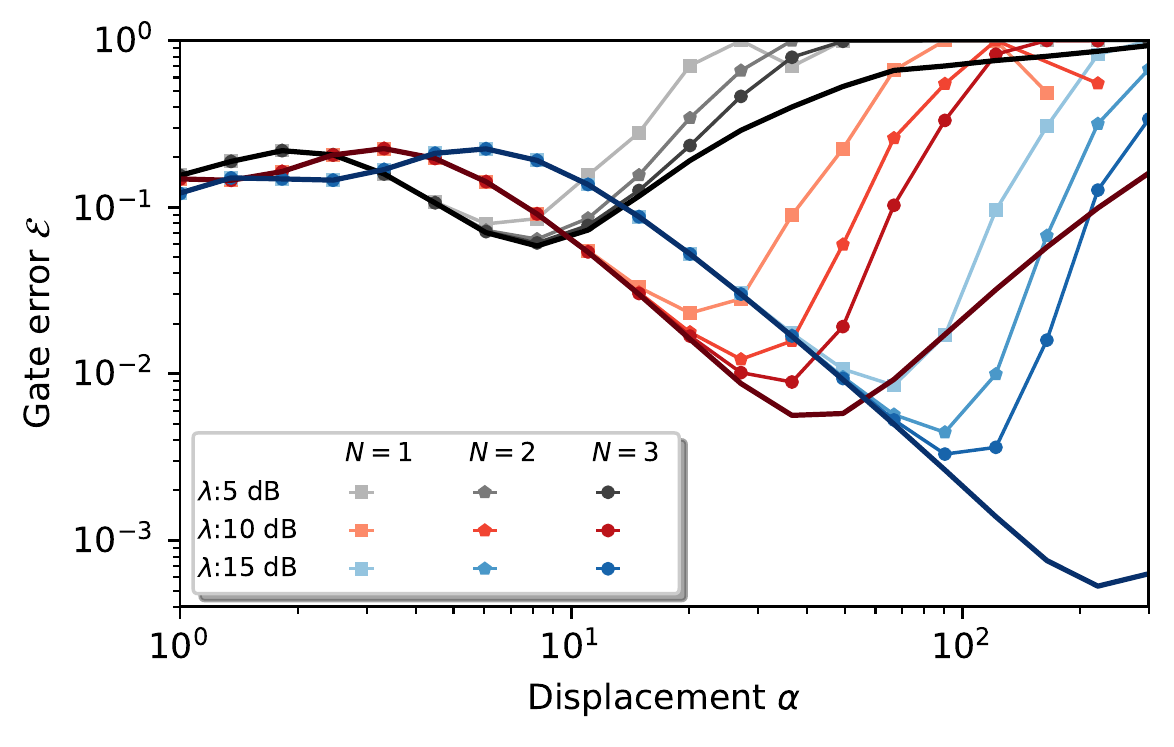}
    \caption{Gate error of the discretized alternative scheme introduced in Appendix~\ref{sec:trotter} (markers), compared to that of the continuous drive scheme that was considered in the main text (solid lines). The gate error is shown as a function of the displacement $\alpha$ with squeezing $\lambda$ being 5 dB (grey), 10 dB (red), and 15 dB (blue). Here, the gate angle is chosen to be $\gamma=0.1$, the input quantum state is chosen to be $\ket{z^+}$ with $\Delta=0.5$, and the lossless $\kappa=0$ case is considered.}
    \label{fig:bch_combined}
\end{figure}

In Fig.~\ref{fig:bch_combined}, we show the gate error of $\hat{U}_\mathrm{discrete}^\mathrm{cubic}(\gamma=0.1,N)$ as a function of the displacement $\alpha$ for various squeezing $\lambda$ and $N$. The figure shows that even $N=1$ can provide reasonably good gate fidelity as long as the displacement is chosen appropriately.

\section{Noise analysis}
\label{sec:noiseanalysis}
In the main text, we analyzed the robustness of our scheme by studying the effect of stochastic phase rotations at the end of the nonlinear medium. In this section, we extend this analysis to other major noise sources in our scheme, by studying the effect of a noisy detuning $\delta$ and drive $\beta$. 

First, we consider the effect of a noisy detuning. Following the approach in the main text, we model this noise by a fixed detuning $\delta= \delta^\mathrm{cubic}+\mathrm{d}\delta$, where $\mathrm{d}\delta$ is an addition detuning due to noise. By substituting this expression of $\delta$ into \eqref{eq:Heffkerr}, we find that the effective Hamiltonian picks up the following additional noise term:
\begin{align}
    \mathrm{d}\delta \hat{a}_{\mathrm{eff}}^\dagger\hat{a}_{\mathrm{eff}}  \approx  3\sqrt{2}\chi\lambda\alpha^3 \frac{\mathrm{d}\delta}{\delta^\mathrm{cubic}} \hat{x}.
    \label{eq:detuning-noise}
\end{align}
Because this term commutes with the cubic phase Hamiltonian $\hat H_{\mathrm{cubic}}$ to leading order in $1/\lambda$, it can be shown that the effect of the noise in the detuning is equivalent to the application of a stochastic $p$-displacement on the output (or input) state of the cubic phase gate. Since the amount of displacement is $3\sqrt{2}\chi\lambda\alpha^3 \frac{\mathrm{d}\delta}{\delta^\mathrm{cubic}} \tau = 6\gamma\alpha^2\lambda^{-2}\frac{\mathrm{d}\delta}{\delta^\mathrm{cubic}}$, the gate error due to this effect scales as $\left(\frac{\mathrm{d}\delta}{\delta^\mathrm{cubic}}\lambda^4\right)^2$. 

Next, we consider the effect of a noisy drive $\beta=\beta^\mathrm{cubic}+\mathrm{d}\beta_\mathrm{x}+\mathrm{i}\mathrm{d}\beta_\mathrm{p}$, where $\beta_\mathrm{x}$ ($\beta_\mathrm{p}$) represents in the in-phase (quadature-phase) noise of the drive. For this noise, the dominant error terms in the effective Hamiltonian are given by
\begin{align}
    \begin{split}
    (\mathrm{d}\beta_\mathrm{x}+\mathrm{i}\mathrm{d}\beta_{\mathrm{p}})\hat{a}_{\mathrm{eff}}^\dagger+(\mathrm{d}\beta_\mathrm{x}-\mathrm{i}\mathrm{d}\beta_\mathrm{p})\hat{a}_{\mathrm{eff}}\approx-2\sqrt{2}\chi\lambda\alpha^3\hat{x}\frac{\mathrm{d}\beta_\mathrm{x}}{\beta^\mathrm{cubic}}.
\end{split}
\end{align}
Thus, we find that the scheme is primarily sensitive to the in-phase noise $\mathrm{d}\beta_\mathrm{x}$. Since this error term takes the same form as \eqref{eq:detuning-noise}, it can be shown that the effect of this noise is once again equivalent to the application of an additional $p$-displacement on the output (or input) state of cubic phase gate. The amount of displacement is given by $4\mathrm{i}\gamma\alpha^2\lambda^{-2}\frac{\mathrm{d}\beta_\mathrm{x}}{\beta^\mathrm{cubic}}$, and the gate error due to this effect scales as $\left(\frac{\mathrm{d}\beta_\mathrm{x}}{\beta^\mathrm{cubic}}\lambda^4\right)^2$. 

\begin{figure}[h]
\begin{center}
\includegraphics[width=0.93\textwidth]{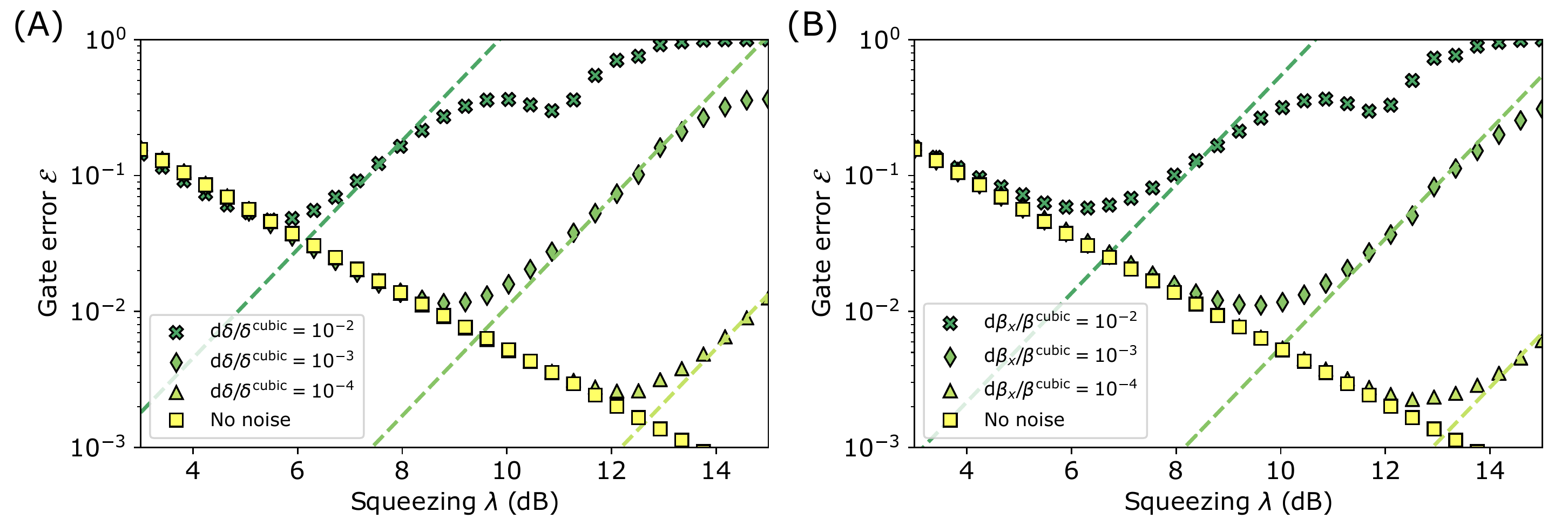}
\caption{Gate error in the presence of a (A) noisy detuning $\delta= \delta^\mathrm{cubic}+\mathrm{d}\delta$ or (B) noisy drive $\beta=\beta^\mathrm{cubic}+\mathrm{d}\beta_x$. For both plots, the input state is the GKP qubit state \eqref{eq:gkp} $\ket{z^+}$ with $\Delta=0.5$, and $\gamma = 0.1$. The dashed lines are guide for eyes showing the expected scaling $\mathcal{E}\sim\lambda^8$.}
\label{fig:paramnoise}
\end{center}
\end{figure}

Fig.~\ref{fig:paramnoise} shows numerical simulations of the gate error in the presence of a noisy detuning or a noisy drive. The results confirm our theoretically expected scaling of $\mathcal{E}\sim\lambda^8$. 

\section{Quantum Kerr solitons}
\label{sec:qsoliton}
In this section, we derive conditions under which optical Kerr solitons can serve as a platform for implementing the cubic phase gate scheme. When a pulse propagates in a nonlinear medium, it is well known that it generally experiences spectral-temporal mode distortions~\cite{Boivin1994,Joneckis1993,Schmidt1998,Blow1991,Leung2009,Brecht2011}. Even in the absence of linear dispersion, the presence of a positive $\chi^{(3)}$ nonlinearity, for example, can induce a phase chirp in the pulse envelope so that the leading end of the pulse has lower frequency. This can be problematic for the construction of a cubic phase gate, because it becomes difficult to assume the single-mode nature of the Kerr interaction, which is essential for our scheme. On the other hand, in classical pulsed nonlinear optics, the introduction of a finite anomalous group velocity dispersion in 1D waveguides enables the formation of Kerr solitons whose envelopes are temporally stable~\cite{Chen2007}. The description of Kerr soliton formation was extended to the quantum setting in Refs.~\cite{Boivin1994,Lai1998, Lai1998b, Drummond1987, Carter1991, Rosenbluh1991,Haus1990}. Following these results, Ref.~\cite{Wright1991,Korolkova2001} analyzed coherent-state excitation of such solitons; when these excitations are large, the system exhibits behavior that can approximately be described by single-mode self-phase modulation (SPM).

Here we briefly review the quantum construction of Kerr solitons, and we adapt the arguments of Ref.~\cite{Wright1991,Korolkova2001} to show that highly-displaced quantum states also undergo single-mode SPM, which is the condition necessary to realize our scheme. In the process, we connect experimentally-relevant parameters to our figure of merit $\chi/\kappa$. Note that, though we used a convention $\hbar=1$ in the main text for simplicity, we write down $\hbar$ explicitly in this section to avoid confusion. Since $\chi/\kappa$ is a unitless quantity, this convention does not lead to any ambiguity.

We consider a 1D nonlinear waveguide along the $z$-direction with positive nonlinear index of refraction $n_2$ and a constant anomalous group velocity dispersion $\partial_k^2\omega$. Using the time-dependent Hartree approximation, Ref.~\cite{Lai1998} shows that one can construct an $n$-photon nonlocal field operator of the form
\begin{align}
    \hat{a}_n=\int h_n(z)\hat{\psi} (z)\mathrm{d}z,
    \quad\text{with}\quad
    h_n(z)=\frac{1}{\sqrt{2v_\mathrm{g}T_n}}\mathrm{sech}\left(\frac{z}{v_\mathrm{g}T_n}\right),
\end{align}
where the local photon-polariton field annihilation operators obey $[\hat{\psi}(z),\hat{\psi}^\dagger(z')]=\delta(z-z')$, and $h_n(z)$ is the normalized $n$-photon wavepacket envelope. Here, $v_\mathrm{g}$ is the group velocity at the carrier frequency $\omega_0$, and $T_n=2\partial_k^2\omega /[(n-1) \hbar\omega_0 v_\mathrm{g}^3\gamma_\mathrm{nl}]$ is the characteristic time scale of the pulse. The effective nonlinear coefficient $\gamma_\mathrm{nl}=\frac{\omega_0 n_2}{c A_\mathrm{eff}}$ is related to the speed of light in vacuum $c$ and the effective cross-section of the waveguide $A_\mathrm{eff}$. Using these field operators $\hat a_n$, we can then construct $n$-photon states which evolve according to
\begin{align}
\label{eq:solitonevolution}
    \ket{\phi_n(t)}=\exp\left[-\frac{\mathrm{i}}{2}\left(-\frac{\hbar  \omega_0v_\mathrm{g}\gamma_\mathrm{nl} n(n-1)}{ 2T_n}\right)t\right]\times\frac{1}{\sqrt{n!}}(\hat{a}_n^\dagger)^n\ket{0}.
\end{align}
Excitations of this form, which propagate with only an intensity-dependent (Kerr) phase shift of its wavepacket, are called quantum Kerr solitons.  Note this solution assumes an instantaneous Kerr interaction, which is a good approximation when the response time of the Kerr effect $T_\mathrm{Kerr}$~\cite{Joneckis1993,Shapiro2006,Boivin1994} in the physical system is much smaller than the characteristic time scale of the pulses (i.e., $T_\mathrm{Kerr}\ll T_n$).  This condition is generally valid in nanophotonic platforms where $T_\mathrm{Kerr}$ is generally on the order of few fs for solid-state materials~\cite{Ziolkowski1993,Agrawal2012,Claude1998,Shtyrkova17}.

\begin{figure}[h]
    \includegraphics[width=0.90\textwidth]{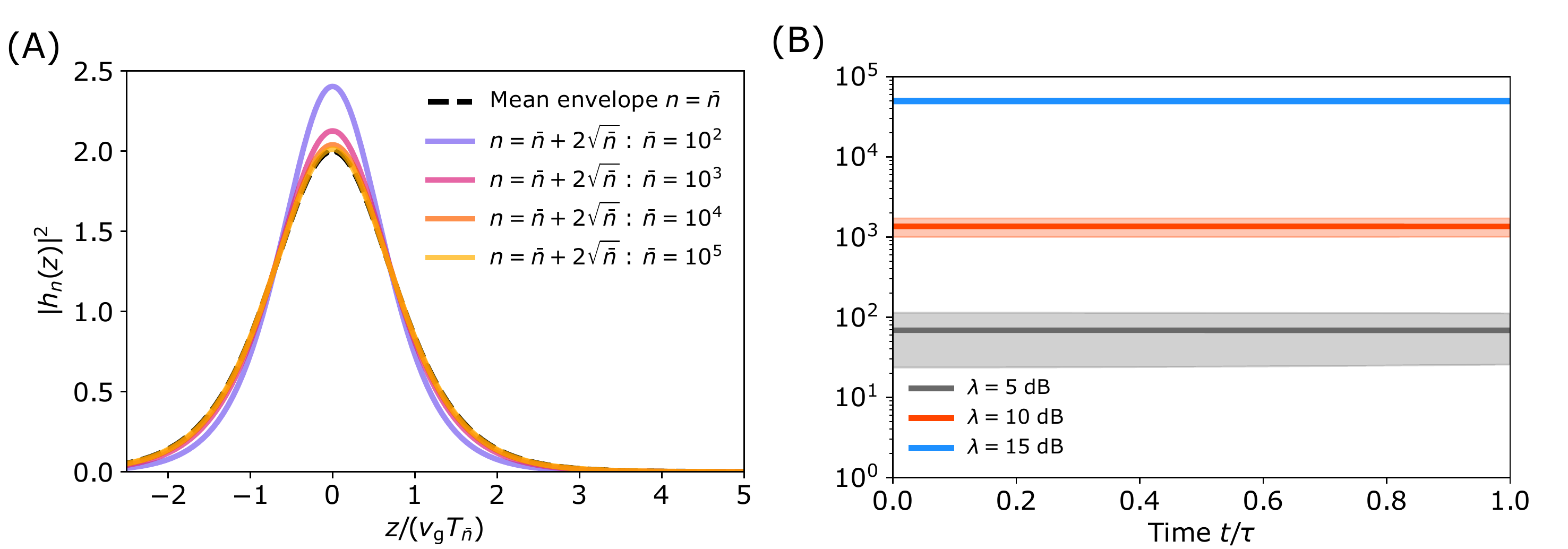}
    \caption{(A) Variation in the wavepacket envelopes $h_n(z)$ for $n$-photon quantum Kerr solitons around a ``mean'' wavepacket $h_{\bar{n}}(z)$ for various $\bar{n}$. The horizontal axis is normalized with respect to the characteristic pulse width $v_{\mathrm{g}}T_{\bar{n}}$ of the mean envelope. (B) Time evolution of the photon number during the operation of a coherent cubic phase gate, showing that the photon number (and hence the soliton wavepacket envelope) does not change appreciably during the gate operation. The solid line indicates the average photon number while the shaded region represents two standard deviations from the average. The simulation is performed using $\gamma=0.1$ on the input state $\ket{z^+}$ with $\Delta=0.5$; the displacement $\alpha$ is set to the value that maximizes the fidelity of the gate for each squeezing parameter $\lambda$.}
    \label{fig:hn}
\end{figure}

Since the excitations described by \eqref{eq:solitonevolution} are definite-photon states, however, coherent-state excitations of these wavepackets necessarily lead to quantum dispersion in the wavepacket. However, we note that for our cubic phase gate, we are primarily concerned with the evolution of quantum states $\ket{\psi} = \hat D(\alpha) \ket{\varphi}$ that are highly displaced, where the excitation in $\ket{\varphi}$ is reasonably small. Thus, if the average excitation of our displaced state is $\bar n \simeq|\alpha|^2$, then following the argument presented in Refs.~\cite{Wright1991, Korolkova2001,Drummond2014}, we can consider the amount of wavepacket spread over a range of photon numbers $\bar{n}-2\sqrt{\bar{n}} < n < \bar{n}+2\sqrt{\bar{n}}$; when $\bar n$ is large, the wavepacket envelopes in this range are almost identical, i.e., $h_n \approx h_{\bar n}$, as shown in Fig.~\ref{fig:hn}(A). Thus, $\hat a_n \approx \hat a_{\bar n}$ in this regime, so if the quantum state is prepared in the wavepacket $h_{\bar{n}}$ (i.e., $\ket{\psi} = \sum_n c_n (\hat{a}_{\bar{n}}^ \dagger)^n\ket{0}$), then its evolution through the medium is described by a simple intensity-dependent phase shift as in \eqref{eq:solitonevolution}, which is equivalent to the action of a single-mode Kerr Hamiltonian
\begin{equation}
\label{eq:bare}
\hat{H}_{\bar{n}}/\hbar=-\frac{\chi}{2}\hat{a}_{\bar{n}}^\dagger{}\hat{a}_{\bar{n}}(\hat{a}_{\bar{n}}^\dagger{}\hat{a}_{\bar{n}}-1)=-\frac{\chi}{2}\hat{a}_{\bar{n}}^\dagger{}^2\hat{a}_{\bar{n}}^2,
\end{equation}
with $\chi = \hbar\omega_0v_\mathrm{g}\gamma_\mathrm{nl}/2T_{\bar{n}}>0$, as desired.

For the construction of a coherent cubic phase gate, an additional linear drive and detuning to the bare Kerr Hamitonian \eqref{eq:bare} has to be introduced. For example, these interactions can be realized by coherently populating an auxiliary waveguide that is evanescently coupled to the main waveguide. This can be formalised as follows: Using the local field annihilation operator for the auxiliary waveguide $\hat{\Psi}(z)$, the evanescent coupling is described by an interaction term $\hat{H}_{\mathrm{int}}/\hbar=\epsilon\int \hat{\Psi}(z)^\dagger\hat{\psi}(z)\,\mathrm{d}z+\mathrm{H.c.}$, where $\epsilon$ is the rate of the coupling. To ensure a single-mode drive, the auxiliary waveguide must be populated with a temporally stable coherent pulse with the same envelope as the signal pulse $h_{\bar{n}}(z)$. Specifically, in the limit where the displacement of the pulse $\xi$ in the auxiliary waveguide is sufficiently large (i.e., $\xi \rightarrow \infty$) and the evanescent coupling is sufficiently weak (i.e., $\epsilon \rightarrow \beta/\xi$) such that $\beta = \epsilon \xi$ is a constant, we can adiabatically eliminate the mode in the auxiliary waveguide through the substitution $\hat{\Psi}(z)\mapsto e^{-\mathrm{i}\delta t}\xi h_{\bar{n}}(z)$, where $\delta$ is the detuning of the auxiliary waveguide mode with respect to the main waveguide mode. This leads to the following single-mode expression of the interaction Hamiltonian: 
\begin{align}
\hat{H}_{\mathrm{int}}/\hbar=e^{\mathrm{i}\delta t}\beta\int  h_{\bar{n}}(z)\hat{\psi}(z)\mathrm{d}z+\mathrm{h.c.}=e^{\mathrm{i}\delta t}\beta\hat{a}_{\bar{n}}+\mathrm{H.c.}.
\end{align}
By moving into the rotating frame via $\hat{a}_{\bar{n}}\mapsto e^{-\mathrm{i}\delta t}\hat{a}_{\bar{n}}$, we arrive at the desired single-mode Hamiltonian
\begin{align}
    \hat{H}_\mathrm{Kerr}/\hbar=-\frac{\chi}{2}\hat{a}_{\bar{n}}^\dagger{}^2\hat{a}_{\bar{n}}^2+\delta \hat{a}_{\bar{n}}^\dagger \hat{a}_{\bar{n}}+\beta(\hat{a}_{\bar{n}}^\dagger+ \hat{a}_{\bar{n}}).
\end{align}

Although the photon number is not generally conserved under a linear drive, this effect is fortunately mostly canceled by the combined effect of the detuning and the Kerr nonlinearity. This is shown in Fig.~\ref{fig:hn}(B), which plots the evolution of average photon number of the state during the prescribed operation of a coherent cubic phase gate. We observe no significant variation in photon number, indicating that the form of the single-mode envelope $h_{\bar{n}}$ set by the mean photon number $\bar{n}$ of the initial state remains valid throughout the sequence.

Finally, we would like to express the figure of merit $\chi/\kappa$ in terms of experimentally relevant parameters. The linear attenuation coefficient  $\alpha_\mathrm{att}$ (in units of \si[power-font=unit]{\dB}[length]${}^{-1}$) to the dissipation rate $\kappa$ via $\kappa=10\alpha_\mathrm{att} v_\mathrm{g}/\log 10 \approx 4.342\alpha_\mathrm{att} v_\mathrm{g}$, while the characteristic time scale $T_n$ is related to the intensity full width at half-maximum $T_\mathrm{FWHM}$ of the sech pulse via $T_\mathrm{FWHM}=(2\arccosh \sqrt{2}) T_{\bar{n}} \approx\num{1.763}T_{\bar{n}}$. Using these relations, the figure of merit is
\begin{align}
    \frac{\chi}{\kappa}\approx\frac{\num{3.83}\hbar\omega_0\gamma_\mathrm{nl}}{T_\mathrm{FWHM}\alpha_\mathrm{att}}.
    \label{eq:fom_app}
\end{align}
We compute the figure of merit $\chi/\kappa$ using experimentally demonstrated values of $\gamma$ and $\alpha$ in nonlinear waveguides assuming $T_\mathrm{FWHM}=\SI{100}{fs}$. Though careful dispersion engineering will be required to form a soliton with the desired length $T_\mathrm{FWHM}$, the calculations suggest that the prospect of realizing our scheme in on-chip photonic architectures is promising in the near future.

\begin{table}[h!]
\begin{center}
\caption{Figure of merit $\chi/\kappa$ computed via \eqref{eq:fom_app} with experimentally demonstrated values of $\gamma_\mathrm{nl}$ and $\alpha_\mathrm{att}$, and with an assumed pulse length of $T_\mathrm{FWHM}=\SI{100}{fs}$.}
   \label{table:materials}
\begin{tabular}{ |p{3cm}||p{2.5cm}|p{2.5cm}|p{2.5cm}|p{2.5cm}|p{1.8cm}|}
    \hline
    Platform &  $\gamma_\mathrm{nl}~(\mathrm{W^{-1}\cdot m^{-1}})$& $\alpha_\mathrm{att}~(\mathrm{dB\cdot m^{-1}})$& Wavelength $(\mathrm{\mu m})$& Fig.\ of Merit $\chi/\kappa$ & Reference\\
    \hline
    Silicon-on-insulator &280 &400 &1.54 & $3.4\times 10^{-6}$&\cite{Clemmen2009}\\
    \hline
    AlGaAs-on-insulator&660 &140 &1.59 & $2.2\times 10^{-5}$ &\cite{Pu2016} \\
    \hline
    Si${}_3$N${}_4$&1 &1 &1.5 & $5.1\times 10^{-6}$ &\cite{Pfeiffer2018} \\
    \hline
   \end{tabular}
   
\end{center}
\end{table}

\twocolumngrid
\bibliography{myfile}
\end{document}